\documentstyle[aps,prd,multicol,epsfig]{revtex}
\begin{document}


\title{Analytic model for galaxy and dark matter clustering}

\author{Uro\v s Seljak}
\address{
Department of Physics, Jadwin Hall, Princeton University,
	Princeton, NJ 08544
}
\date{January 2000}

\maketitle
\begin{abstract}
We investigate an analytic model to compute nonlinear power spectrum 
of dark matter, galaxies and their cross-correlation. The model 
is based on Press-Schechter halos, which cluster and have 
realistic dark matter profiles. 
The total power spectrum is a sum of two contributions, one from 
correlations betwen the halos and one from correlations within the same
halo. We show that such a model 
can give dark matter power spectra which match well with the results of 
N-body simulations, provided that concentration parameter decreases 
with the halo mass. 

Galaxy power spectrum differs from dark matter power spectrum 
because pair weighted number of galaxies does not scale with the halo
mass and because most halos harbor 
a central galaxy. If the pair weighted
number of galaxies increases less rapidly than the halo mass, 
as predicted 
by theoretical models and observed in clusters, then the resulting power 
spectrum becomes a power law with the slope closed to the observed 
over several orders of magnitude in scale. Such a model also predicts
later onset of nonlinear clustering compared to the dark matter, 
which is needed to reconcile the CDM models
with the data. Generic prediction of
this model is that bias is scale dependent and nonmonotonic.  
This is particularly important for red or elliptical galaxies, which
are preferentially found in larger mass halos and for which bias in 
power spectrum may be scale dependent even on large scales.

Our predictions for galaxy-dark matter correlations, which can be 
observed through the galaxy-galaxy lensing, show that these cannot be 
interpreted simply as an average halo profile of a typical galaxy,
because different halo masses dominate at different scales and 
because larger halos host more than one galaxy. We 
compute predictions for the cross-correlation coefficient 
as a function of scale and discuss the prospects of using 
cross-correlations in combination with galaxy clustering to 
determine the dark matter power spectrum.

\end{abstract}
\newcommand{\tl}{\tilde}
\newcommand{\bm}{\boldmath}
\newcommand{\cut}{\rm cut}
\newcommand{\mnras}{Mon. Not. Roy. Astron. Soc.}
\def\bi#1{\hbox{\boldmath{$#1$}}}
\def\sun{\hbox{$\odot$}}

\section{Introduction}
Correlations in dark matter contain a wealth of information 
about cosmological parameters. Their power spectrum is sensitive
to parameters such as matter density, Hubble constant, primordial 
power spectrum slope and amplitude, massive neutrinos, baryon 
density etc. Determining the linear power spectrum of dark matter 
is one of the 
main goals of modern cosmology. There are several complications 
that prevent us at present from reaching this goal. First, 
on small scales the linear power spectrum is modified
by nonlinear evolution which enhances its amplitude over the 
linear spectrum. It is important to understand this process, so 
that one can predict the relation between the two. 
This is necessary both to reconstruct the linear spectrum from 
a measured nonlinear one and to verify whether there are other
mechanisms besides gravity that modify the clustering of dark matter 
on small scales. 
Examples of such are baryonic feedback effects on dark matter \cite{DS} or
nongravitational interactions between dark matter 
particles \cite{SS99}. Second, it is difficult to observe correlations in 
dark matter directly. Direct tracers such as peculiar velocity flows
or weak lensing still suffer from low statistics and poorly understood
systematics. Instead it is much easier to
observe correlations between galaxies \cite{P97} or 
correlations between galaxies and dark matter \cite{qsogal}.
While these are related to the dark matter 
correlations, the relation may not be simple. The goal of this 
paper is to address both issues with 
a model that is simple enough to allow analytic calculations without 
the use of N-body simulations, yet sufficiently accurate
to be useful for predicting galaxy and dark matter power spectrum.

Our approach to dark matter clustering 
is based on the Press \& Schechter model \cite{PS74}. In 
this picture at any given time 
all the matter in the universe is divided into virialized
halos. These halos are correlated and have some internal density 
profile, which can be a function of halo mass. By specifying 
the halo mass function, their clustering strength and their
halo profile we can determine the dark matter correlation function. 
The formalism for correlations inside halos has been 
developed by \cite{MS} and applied to power law 
halos \cite{SJ}. We generalize this approach by including 
the correlations between halos and by using more realistic 
non-power law halo profiles whose shape may 
depends on the halo mass \cite{NFW}.
We show in this paper that such a generalized model can provide 
very good agreement with results of numerical simulations over 
a wide range of scales \cite{ScoSh}.

The central question in extracting dark matter power spectrum from 
that of the galaxies is how well galaxies trace dark matter, the 
issue of bias. 
This has been addressed theoretically 
both with hydrodynamic \cite{bias} and semi-analytic methods 
\cite{Benson,Kauffmann}. The fact that the galaxy correlation function is 
a power law over several decades in scale, while power spectra in 
CDM models do not show such behaviour, already indicates that the bias is scale 
dependent. Moreover, galaxies come in different types and observational
data show that they can be biased relative to one another \cite{biasobs}.
In our modelling of galaxy correlations we introduce two new functions, 
the mean number and the mean pair weighted number 
of galaxies inside the halo as a function of the halo mass.
The importance of these has recently been emphasized in the 
context of pairwise velocity measurements \cite{JingMoBo98,Benson99b} 
and galaxy clustering \cite{Benson}. These play a key role in 
understanding the relation between galaxy and dark matter clustering.
We explore the predictions for different choices of these relations 
and compare them to the results of semi-analytic models. 

Galaxy-dark matter correlations can provide additional information 
on the clustering of galaxies and dark matter and the relation between them. 
Such correlations have been observed through gravitational lensing 
effects, for example using galaxy-galaxy lensing or 
correlations between foreground and background populations \cite{qsogal}. 
Such measurements are often interpreted either 
in terms of an averaged density profile of a halo \cite{brainerd} 
or in terms of a constant bias model \cite{waerbeke}. 
We discuss the applicability of these models and how they can be 
generalized to take into account effects such as 
broad range of halo masses and multiple galaxies inside halos. 

\section{Dark matter power spectrum}

The halo model for power spectrum assumes all the matter is in a
form of isolated halos with a well defined mass $M$ and halo profile 
$\rho(r,M)$. The halo profile is defined to be an average over all 
halos of a given mass and does not necessarily assume all halos have
the same profile. 
The mass is determined by the total mass within the virial 
radius $r_v$, 
defined to be the radius where the mean density within it is $\delta_{\rm vir}$ times 
the mean density of the universe.
Throughout the paper we will use 
$\Lambda CDM$ model
with $\Omega_m=0.3$, $\Omega_{\Lambda}=0.7$,
normalized to $\sigma_8=0.9$ today. 
For this model $\delta_{\rm vir} \sim 340$, although 
we will also use $\delta_{\rm vir} \sim 200$ (the value
for Einstein-de Sitter universe) for 
consistency with the results of some of the N-body 
simulations.
The halo profile is spherically averaged and 
assumed to depend only on the mass of the halo. 
We will model the halo density profile in 
the form 
\begin{equation}
\rho(r)={\rho_s \over (r/r_s)^{-\alpha}(1+r/r_s)^{3+\alpha}}.
\label{rho}
\end{equation}
This model assumes that the profile shape is 
universal in units of scale radius $r_s$, while its characteristic density 
$\rho_s$ at $r_s$ or concentration $c=r_v/r_s$ may depend on the halo mass. 
The halo profile is assumed to go as
$r^{-3}$ in the outer parts and as $r^{\alpha}$
in the inner parts, with the transition between the two at $r_s$.
The outer slope is fixed by the results of N-body simulations which 
generally agree in this regime.  
An example of such a profile is $\alpha=-1$ 
\cite{NFW,Huss,Tormen}.  
Other models have however been proposed 
with 
$\alpha=-1.5$ \cite{Moore,Fukushige} or even $\alpha>-1$ \cite{Kravtsov}. 
In principle $\alpha$ could also be a 
function of mass scale and may steepen towards 
smaller mass halos 
with $\alpha \sim -1$ for cluster halos and 
$\alpha \sim -1.5$ for galactic halos
\cite{JingSuto99}. Similarly, concentration $c$
may depend on the mass and different authors find somewhat different 
dependence \cite{NFW,Moore,Bullock99}. 
We will explore how variations
in the profile and concentration affect the power spectrum.
Instead of $r_s$ we use the concentration parameter $c=r_v/r_s$
as a free parameter. Note that $r_v$ is related to $M$ via 
$M=4\pi/3r_v^3\delta_{\rm vir}\bar{\rho}$. Similarly 
we can eliminate $\rho_s$ and 
describe the halo only in terms of its virial mass $M$ and 
concentration $c$, 
because the integral over the halo density profile (equation \ref{rho}) 
must equal the halo mass.

For a complete description we need 
the halo mass function $dn / dM$, describing the number density of halos 
as a function of mass. It can be written as
\begin{equation}
{dn \over dM} dM={\bar{\rho} \over M}f(\nu)d\nu,
\end{equation}
where $\bar{\rho}$ is the mean matter density of the universe. We
introduced
function $f(\nu)$, which can be 
expressed in units in which it has a universal form independent of the
power spectrum or redshift if written
as a function of peak height 
\begin{equation}
\nu=[\delta_c(z)/\sigma(M)]^2. 
\label{nu}
\end{equation}
Here $\delta_c$ is the value of a spherical overdensity at which it
collapses at $z$ ($\delta_c=1.68$ for Einstein-de Sitter model) and
$\sigma(M)$ is the rms fluctuation in spheres that contain on average
mass $M$ at initial time, extrapolated using linear theory to $z$.
The form proposed by Press \& Schechter (PS) \cite{PS74} is  $\nu f(\nu)=
(\nu/2\pi)^{1/2}e^{-\nu/2}$. This has been shown to overpredict the 
halo abundance by a factor of 2 
at intermediate masses below nonlinear mass scale $M_*$ 
\cite{Somm,ShethTormen99}. 
A modified version of this form
that fits better the N-body simulations is given by Sheth \& Tormen (ST)
\cite{ShethTormen99} 
\begin{equation}
\nu f(\nu)=A(1+ \nu'^{-p})\nu'^{1/2} e^{-\nu'/2},
\end{equation}
where $\nu'=a\nu$ with $a=0,707$ and $p=0.3$ as the best fitted values,
which gives $\nu f(\nu) \propto \nu^{0.2}$ for small $\nu$. 
PS expression 
corresponds to $a=1$, $p=0$ giving $\nu f(\nu) \propto \nu^{0.5}$ for small 
$\nu$. The
constant $A$ is determined by mass conservation, 
requiring that the integral over the mass 
function times the mass gives the mean density 
\begin{equation}
{1 \over \bar{\rho}}\int {dn \over dM}M dM=\int f(\nu)d\nu=1.
\label{rhobar}
\end{equation}
Note that we can still apply this equation even if some dark matter is not bound
to any halo. In this case the mass function has a nonvanishing contribution 
in the limit $M\rightarrow 0$. 

The correlation function consists of two terms. 
On large scales the halos are correlated with each other. We assume 
the halo-halo correlation function follows the linear correlation function. 
Its amplitude depends on the bias for each halo. Halos more massive 
than the nonlinear mass scale $M_*$ are more strongly clustered than the
matter, while those with masses below $M_*$ are less strongly clustered
than the matter. A simple halo biasing scheme has been given by 
\cite{ColeKaiser89,MoWhite96} and generalized to the ST mass function 
by \cite{ShethTormen99}
\begin{equation}
b(\nu)=1+{\nu -1 \over \delta_c}+{2p \over \delta_c(1+\nu'^p)}.
\label{b}
\end{equation}

Since halos are not pointlike we need to convolve the halo-halo 
correlation function with the halo profiles of both halos to 
obtain the dark matter correlation function.
The expressions simplify significantly in Fourier space, where convolution 
becomes a multiplication with the Fourier transform of
the halo profile 
\begin{equation}
\tilde{\rho}(k,M) = \int 4\pi r^2 dr \rho(r,M){\sin(kr) \over kr}.
\end{equation}
Note that this is normalized so that $\tilde{\rho}(0,M)=M$.
It is convenient to renormalize it to unity by introducing 
a new variable $y(k,M)=\tilde{\rho}(k,M)/M$, so that 
$y(0,M)=1$ and $y(k>0,M)<1$. The mass of the halo rapidly increases as 
$r^{3-\alpha}$ up to $r=r_s$, but increases only logarithmically 
between $r_s$ and $r_v$ if the outer profile is $\rho(r) \propto r^{-3}$. 
The dominant contribution to the mass therefore comes from radii around $r_s$. 
For $kr_s \ll 1$
we have $y \approx  1$. At $k r_s\sim 1$ there is a transition
and $y$ begins to decrease with $k$, 
so that for $kr_s \gg 1$ we have $y(k,M) 
\propto (kr_s)^{-(3+\alpha)}$. 

\begin{figure}
\centerline{\psfig{file=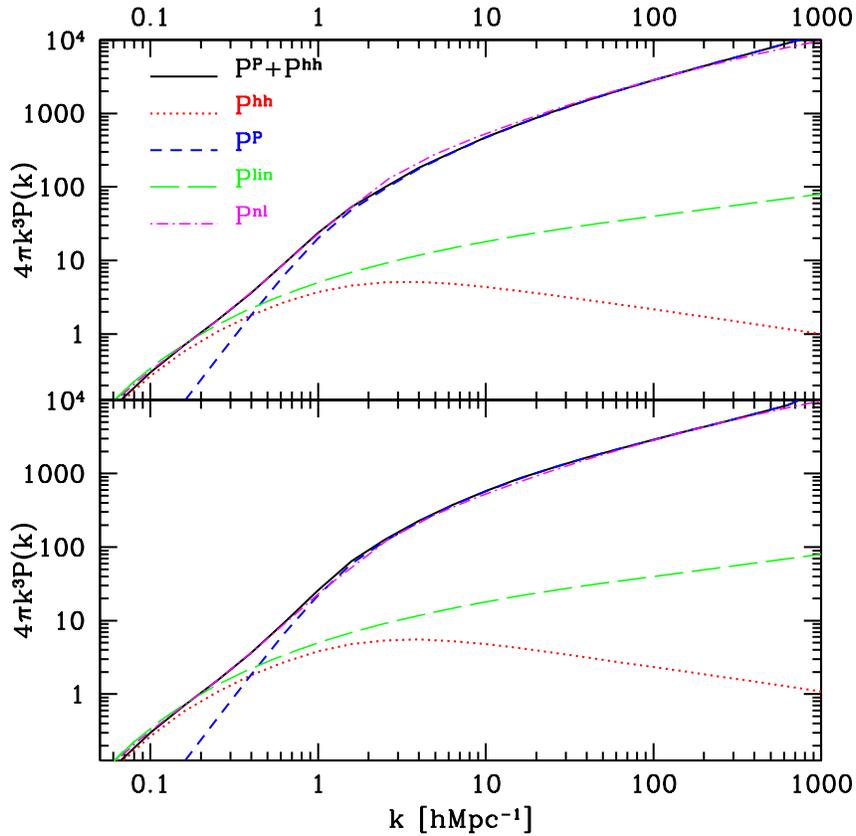,width=4.5in}}
\caption{
Comparison between the power spectrum predicted with our model
and the PD nonlinear power spectrum for $\Lambda CDM$ model. 
Also shown are the linear 
power spectrum and the two individual contributions, $P^P$ and 
$P^{hh}$. Top is the $\alpha=-1.5$ profile, bottom is $\alpha=-1$.
Other parameters are given in the text.
}
\label{fig1}
\end{figure}

Because the expressions simplify significantly in Fourier space 
we will in the following only describe the power spectrum analysis.
The halo-halo term
is given by the integral over their mass function with the appropriate 
bias and the halo profile transform, 
\begin{equation}
P^{hh}_{\rm dm}(k)=P_{\rm lin}(k)\left[\int
f(\nu)d\nu  b(\nu)y[k,M(\nu)]\right]^2, 
\label{hh}
\end{equation}
where $P_{\rm lin}(k)$ is the linear power spectrum and
$M$ is related to $\nu$ via equation \ref{nu} using the 
relation between $\sigma^2(M)=4\pi \int P_{\rm lin}(k)W_R(k)k^2 dk$ 
and $M=4\pi R^3 \bar{\rho}/3$, where $W_R(k)$ is the Fourier transform 
of the top hat window with radius $R$.
This gives $M \propto \nu^{3/(n+3)}$, where $n$ is the slope of the 
linear power spectrum at scale $k \sim R^{-1}$.
We can also define the nonlinear mass scale $M_*$ where $\nu=1$.
Note that on galaxy and smaller scales $n<-2$ and the relation between 
$M$ and $\nu$ is very steep, $M\propto \nu^{\gamma}$ with $\gamma>3$.
The requirement that on large scales ($k \rightarrow 0$, $ y \sim 1$) 
the power spectrum reduces to the 
linear power spectrum imposes a nontrivial 
constraint on the bias distribution,
\begin{equation}
\int f(\nu)d\nu b(\nu) =1.
\label{ndm}
\end{equation}
This implies that if halos are biased ($b>1$) for $M>M_*$ 
at least some of the halos with $M<M_*$
must be antibiased ($b<1$) to satisfy this constraint.
Most of the bias descriptions in the literature  
satisfy this constraint to within a few percent.  
The halo-halo term follows the linear power spectrum on 
large scales and drops below it on scales where finite 
extent of the halos become important (ie where $y(k,M)<1$).
This term is shown in figure \ref{fig1} and as 
expected is dominant on large scales.

In addition to the halo-halo correlation term there are also correlations 
between dark matter particles 
within the same halos. These are expected to dominate on small 
scales. We denote this the Poisson term, which is given by
\begin{equation}
P^P_{\rm dm}(k)= {1 \over (2\pi)^3}\int f(\nu)d\nu {M(\nu) \over \bar{\rho}}
|y[k,M(\nu)]|^2,
\label{pdm}
\end{equation}
The main  
difference between this term and the halo-halo term in equation \ref{hh}
is that we have an additional mass weighting $M/\bar{\rho}$. 
This makes the dominant contribution to this term to come from 
the higher mass halos relative to the halo-halo term. 
On large scales  
($k \rightarrow 0$, $y \sim 1$)
the Poisson term is independent of $k$ and behaves as white noise. 
It increases with $k$ more rapidly than the 
halo-halo term, which scales as the linear power spectrum (figure \ref{fig1}). 
The Poisson term declines below the white noise on small 
scales where the effects of the halo profile become important.   

The total power spectrum is the sum of the two contributions,
\begin{equation}
P_{\rm dm}(k)=P^{hh}_{\rm dm}(k)+P^P_{\rm dm}(k).
\end{equation}
To complete the calculation we need to model the dependence of $c$ on $M$. 
We will parametrize it as 
\begin{equation}
c=c_0(M/M_*)^{\beta}.
\end{equation}
Typical values for $c_0$ are around 10 at the nonlinear mass
scale for $\alpha=-1$ profile \cite{NFW,Bullock99}
and about a third lower for $\alpha=-1.5$
profile \cite{Moore}. Numerical studies also show that 
the concentration decreases slowly with the halo mass, making 
$\beta$ negative.

Figure \ref{fig1} shows the individual contributions and the 
sum in comparison to the linear power spectrum and the nonlinear 
prediction from \cite{PD97} (PD). 
In top of the figure we used 
$\alpha=-1.5$ and $c(M)=6(M/M_*)^{-0.15}$. The latter fits
the concentration mass dependence given in \cite{Moore}. 
Note that 
for consistency with \cite{Moore} 
we use $\delta_{\rm vir}=200$ in this case as opposed to 
$\delta_{\rm vir}=340$.
In bottom of the figure we used the
ST mass function and $\alpha=-1$  with 
$c(M)=10(M/M_*)^{-0.2}$, which is somewhat steeper than numerical 
studies predict \cite{Bullock99} as discussed below. 
The agreement in both cases
is quite remarkable given the simple nature of the model. It correctly 
predicts the transition between the linear and nonlinear power spectrum, as
well as reproduces well the slope at higher values of $k$. 
This shows that 
given a suitable choice of $c(M)$ both models can reproduce the 
nonlinear power spectrum. Conversely, the slope of 
the power spectrum at high $k$ 
is not directly determined by the inner slope of dark matter
profiles, at least if the inner profiles are shallower than $\alpha=-1.5$.

In the case of $\alpha=-1$ profile
the best fitted value for $c_0=10$ agrees 
well with \cite{Bullock99,NFW}, while $\beta \sim -0.2$ is somewhat 
lower than  $\beta=-0.07$ \cite{NFW} and $\beta=-0.13$ \cite{Bullock99}. 
If one adopts such shallow dependence of $c(M)$ with $\beta \sim -0.1$ 
then for $k>10h$Mpc$^{-1}$ the predictions of the model
are systematically below the PD model. Before concluding that this
is caused by the 
galactic halos not being sufficiently compact we must investigate
the possibility that the mass function is underestimated at small masses. 
Replacing ST with PS does not significantly affect the results.
However, both PS and ST
assume that each mass element belongs to only one halo, counting only 
the isolated halos. 
This is certainly a valid description on large scales, where the total halo 
mass determines the white noise amplitude of the power spectrum. On 
small scales the clumpiness caused by subhalos within the halos may
become important.
Recent numerical simulations have in fact shown that most of the 
small halos that merge into larger ones are not immediately 
destroyed, but stay around for some time until they are finally merged
on the dynamical friction time scale \cite{TDS98,Moore,Kravtsov}. 
In such a case a given mass particle
can be part of more than one halo at any given time. Because
on very small scales the correlation function is dominated by the small 
halos it may make a difference whether the mass is 
smoothly distributed within the halos or some fraction of it is in 
the subhalos. 
However, the contribution to the total mass of the
halo coming from the subhalos is below 10\% \cite{Kravtsov,Tormen}.
Recently, the 
mass function for subhalos from high 
resolution simulations was determined and it was
shown that it is an order of magnitude below the one for 
isolated halos \cite{Sigad00}. One may conclude
therefore that subhalos do not affect the mass function significantly 
and cannot resurrect $\beta >-0.13$, $\alpha=-1$ model. 

Steepening the halo profile or changing concentration
can both increase the power spectrum to 
agree better with N-body simulations (figure \ref{fig1}).
This is because to increase the power on small scales one has to increase 
the amount of mass contained within a given radius. This can 
be achieved either by 
making the inner profile steeper than $\alpha=-1$ or making
the concentration parameter larger towards the smaller mass
halos. 
The change in the slope would support the 
results in \cite{Moore,Fukushige}, 
where the universal profile 
has the inner slope close to $\alpha \sim -1.5$, 
or in \cite{JingSuto99},
where the profile is not universal and steepens as the halo mass is 
decreased, so that the inner slope changes from $\alpha \sim 
-1$ on the cluster scales
to $\alpha \sim -1.5$ on the galactic scales. 
If the inner slope of the 
halo profile is $\alpha \sim -1$ the concentration has a stronger mass
dependence than in \cite{NFW,Bullock99}, although the 
discrepancy is not large. As shown in 
figure \ref{fig1} both models can fit the nonlinear power spectrum on small 
scales remarkably well. 

\begin{figure}
\centerline{\psfig{file=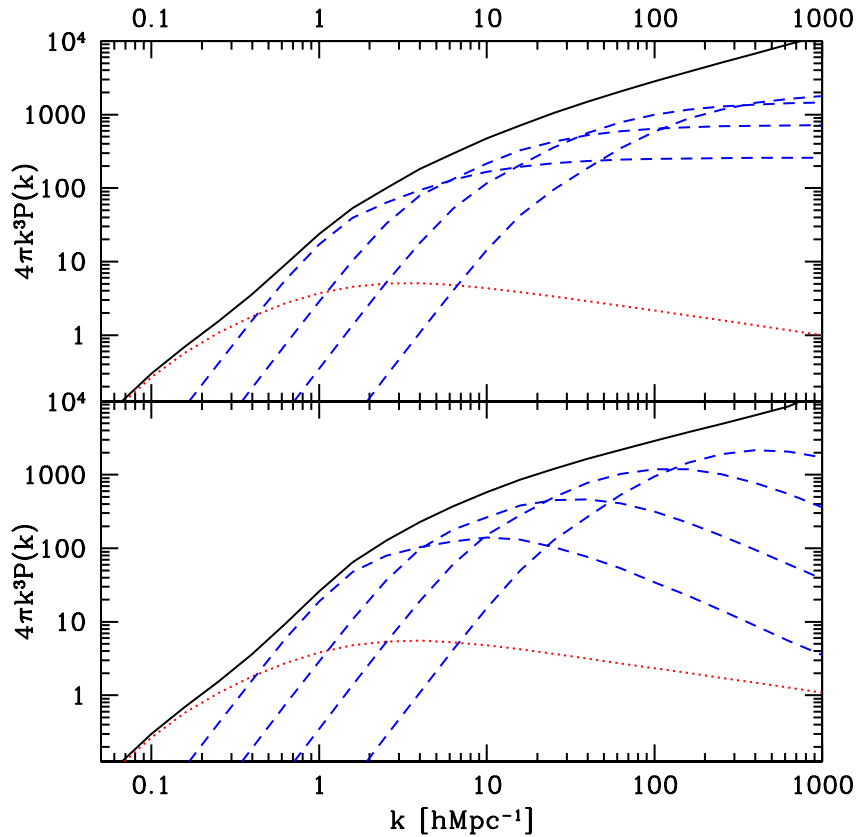,width=4.5in}}
\caption{
Contribution to the $P^P(k)$ from different halo mass intervals for 
the two models in figure \ref{fig1}. 
Short dashed lines from left to right are $M>10^{14}h^{-1}M_{\sun}$, 
$10^{14}h^{-1}M_{\sun}>M>10^{13}h^{-1}M_{\sun}$,
$10^{13}h^{-1}M_{\sun}>M>10^{12}h^{-1}M_{\sun}$ and
$10^{12}h^{-1}M_{\sun}>M>10^{11}h^{-1}M_{\sun}$.
Solid line is the total $P(k)$, 
dotted the correlated term $P^{hh}(k)$.
}
\label{fig2}
\end{figure}

Further insight into the 
relation between the halos and the dark matter power spectrum 
can be obtained by investigating the contribution to the power spectrum 
from different mass intervals. This is shown in figure \ref{fig2} for 
the Poisson term, using the two models from figure \ref{fig1}. 
On large scales the Poisson term is dominated by very 
massive clusters with $M>10^{14}h^{-1}M_{\sun}$. These halos dominate 
the nonlinear clustering on 
scales around and below $k < 1 $hMpc$^{-1}$. On smaller
scales the contribution from large clusters is suppressed  
because $y(k,M)$ begins to decrease from unity at
$ k \sim r_c^{-1} \propto c(M)M^{-1/3} \propto M^{-0.5}$. 
This occurs at lower $k$ for the higher 
mass halos. As a result around
$k \sim 10h$Mpc$^{-1}$ the
halos with $10^{14}h^{-1}M_{\sun}>M>10^{13}h^{-1}M_{\sun}$
dominate, while around $k \sim 100 h$Mpc$^{-1}$ the
halos with $10^{13}h^{-1}M_{\sun}>M>10^{12}h^{-1}M_{\sun}$
dominate.
Note again that the inner slope plays a 
subdominant role in determining the amplitude of the power spectrum. Even if
for a steeper slope the power spectrum from a given mass interval is 
decreasing less rapidly (for example for $\alpha=-1.5$ it is 
asymptotically flat as opposed to decreasing as $k^{-1}$ for $\alpha=-1$),
when this becomes important the
smaller mass halos have already taken over as a 
dominant contribution to the power spectrum.
The nonlinear power spectrum therefore does not reflect the 
inner slope of the halo profile, but rather 
the halo mass function and 
the radius at which the mass enclosed
within this radius begins to deviate significantly from the total halo mass. 
In both models the halos with $M>10^{11}h^{-1}M_{\sun}$ dominate the power 
spectrum for $k<100h$Mpc$^{-1}$. Any modifications in the linear power 
spectrum on mass scales below $M \sim 10^{11}h^{-1}M_{\sun}$ \cite{LK} would 
therefore show up in the dark matter 
correlation function only on kiloparsec scales and below. 

It is interesting to explore in more detail the quasi-linear regime, 
where $P^P(k) \sim {\rm const}$. This approximation is valid up to
$4\pi k^3P(k) \sim 10$ or $k \sim 1h$Mpc$^{-1}$. 
On scales larger than these the power spectrum can be approximated as 
a sum of a linear power spectrum and a constant term, whose amplitude 
is given as an integral over the mass function (equation \ref{pdm} with 
$y=1$). 
From figure \ref{fig2} one can see that the amplitude of this 
integral is dominated by the massive halos, $M>10^{14} h^{-1}M_{\sun}$. 
It is important to emphasize that this amplitude depends only 
on the integral over the power spectrum and not on the details of
the power spectrum itself. Even if there are sharp features in the linear 
power spectrum, such as 
for example baryonic wiggles \cite{TE}, these would not show 
up as features in the quasi-linear power spectrum. Instead, they would 
be integrated over into a single number, corresponding to the mass
weighted integral over the mass function (equation \ref{pdm}). 
This argument is in agreement with the results of N-body simulations
\cite{MWP99} which indeed show that any baryonic features are erased
in the nonlinear regime. This suggests that while the PD model breaks down for 
such spectra, our model could also be applied in such a case. This 
also applies to the spectra with truncated power on small scales 
\cite{White}. We plan to investigate this further in the future.

\section{Galaxy power spectrum}

We now apply the above developed model to the galaxies. 
We assume all the galaxies form in halos, which is a reasonable 
assumption given that only very dense enviroments which have
undergone nonlinear collapse allow the gas to cool and to form stars.
The key new parameters we introduce are the mean number of galaxies 
per halo as a function of halo 
mass, $\langle N \rangle (M)$, and the mean pair weighted number of galaxies
per halo, $\langle N(N-1)\rangle^{1/2}(M)$. Just as in the case of dark matter
these functions are well defined even if the assumption that 
the statistical properties of galaxy population
depend only on the halo mass and not on its enviroment is not satisfied
\cite{bias}, as long as the averaging is performed over all possible 
enviroments. The resulting power spectrum on small scales where the Poisson 
term dominates is independent of this assumption. On large scales where correlations 
between the halos are important violation of this assumption may lead to a change
in the strength of the halo-halo term.

We furthermore assume that there each halo
has a galaxy at its center, while the rest of the 
galaxies in the halos are distributed 
in the same way as the dark matter, so $y(k,M)$ remains unchanged. 
This is only the simplest model and one could easily generalize 
it to profiles that differ from the dark matter.
Any such complications are 
important on small scales, while on large scales ($k<1h$Mpc$^{-1}$) all that is 
relevant is the total number of galaxies inside the halo. 
The normalization equation \ref{ndm} becomes
\begin{equation}
 \int {\langle N \rangle \over M} f(\nu)d\nu={\bar{n} \over \bar{\rho}},
\label{ngal}
\end{equation}
where $\bar{n}$ is the mean density of galaxies in the sample. 

The halo-halo correlation term is given by 
\begin{equation}
P^{hh}_{\rm gg}(k)=P_{\rm lin}(k) \left[{\bar{\rho}\over \bar{n}}\int
f(\nu)d\nu  {\langle N \rangle \over M}b(\nu)y(k,M)\right]^2. 
\label{ghh}
\end{equation}
This should be modified somewhat because the central galaxy does 
not contribute a $y(k,M)$ term, but 
this is only important on small scales where the halo-halo term is 
negligible. 
On large scales where $y\sim 1$ this term 
gives the constant bias model 
\begin{equation}
P^{hh}_{\rm gg}(k)=\langle b\rangle ^2 P_{\rm lin}(k), 
\end{equation}
where the mean galaxy bias $\langle b\rangle$ is given by 
\begin{equation}
\langle b\rangle={\bar{\rho}\over \bar{n}}\int
f(\nu)d\nu  {\langle N \rangle \over M}b(\nu).
\end{equation}

The Poisson term is given by 
\begin{equation}
P^P_{\rm dm}(k)= {\bar{\rho}^2 \over (2\pi)^3 \bar{n}^2}
\int {M \over \bar{\rho}}f(\nu)d\nu {\langle N(N-1)\rangle \over M^2}
|y(k,M)|^p.
\label{gp}
\end{equation}
We use the approximation with $p=2$
if  $\langle N(N-1)\rangle > 1$, because in the limit
the number of pairs is large 
it is dominated 
by the halo galaxies, and $p=1$ if $\langle N(N-1)\rangle <1$, 
because in the opposite limit 
the number of pairs in this case is dominated by 
the central galaxy paired with a halo galaxy. Following the
usual convention \cite{Peebles}
we use $\langle N(N-1)\rangle$ instead of $\langle N^2\rangle$, since
we subtract out the shot noise term arising from the discrete
nature of galaxies (such a term does not depend on the halo profile 
$y(k,M)$). 
Comparing equations \ref{hh} and \ref{pdm} with equations \ref{ghh} and
\ref{gp} we see that there is no difference between the two only if
$\langle N \rangle/M$ and $\langle N(N-1)\rangle^{1/2}/M$ are 
independent of $M$, there are many galaxies per halo and the galaxies
are distributed as the dark matter within the halo. 
For such conditions the power spectrum of galaxies is 
identical to the power spectrum of dark matter. 

\begin{figure}
\centerline{\psfig{file=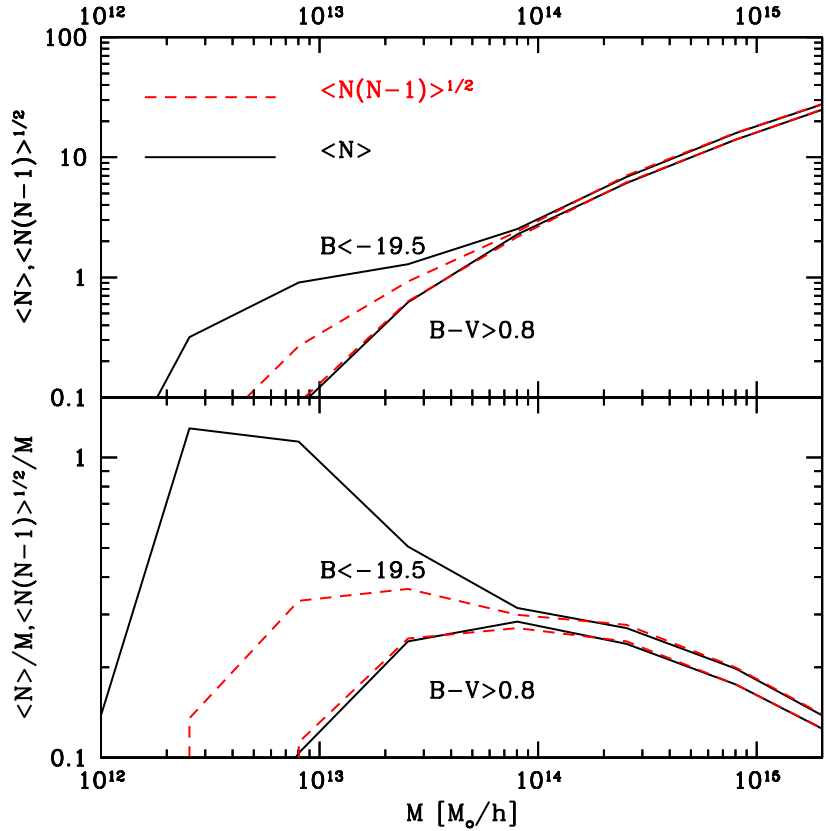,width=4.5in}}
\caption{Top figure shows
$\langle N(N-1)\rangle^{1/2}$ and $\langle N\rangle$ versus $M$
for galaxies selected by absolute magnitude $M_B<-19.5$ (upper curves)
and color $M_B<-19.5$, $M_B-M_V>0.8$ (lower curves) from semi-analytic models. 
Bottom figure shows  the same functions divided by $Mh/10^{13}M_{\sun}$.
}
\label{fig3}
\end{figure}

To test the model above we use 
semi-analytic models of galaxy formation developed in \cite{Kauffmann}. 
These models use N-body simulations to identify the halos and their 
progenitors. Gas is assumed to follow dark matter initially so that 
it heats up during the collapse to the virial temperature of the halo. 
Because of the high density it can efficiently cool and subsequently 
concentrate at the center of the halo. Stars are formed from this cold 
gas on the dynamical time scale. The parametrized star formation 
efficiency and the stellar population synthesis models are used to assign 
magnitudes in different color bands 
to the formed galaxies. The small halos with 
galaxies in them subsequently 
merge into larger halos and exist as individual galaxies until they 
merge with the central galaxy on the dynamical friction time scale. 
The output of 
these models is a catalog of halos and their masses.
For each halo the output consists of a list of 
galaxies, their positions and
luminosities in different bands.
From such a catalog one can reconstruct the 3-d distribution of 
galaxies and dark matter, as well as
$\langle N \rangle$ and $\langle N(N-1)\rangle^{1/2}$
averaged over a given range of halo masses for any desired galaxy 
selection criterion. The goal of our comparison is to compare
the galaxy power spectrum predicted from our model using 
$\langle N \rangle(M)$ and $\langle N(N-1)\rangle^{1/2}(M)$
from semi-analytic models to the galaxy power spectrum obtained 
directly from these models. This is a meaningful comparison even 
if semi-analytic models do not correctly describe the nature.
If we determine that the model contains all the necessary 
ingredients to predict the galaxy correlations we can then try to obtain
these ingredients by other means, either through direct observations 
or better modelling. This can also be applied in the other direction: 
from observations of galaxy power spectrum (and galaxy-dark matter 
power spectrum discussed in the next section) we can determine the 
ingredients of our model, which must be satisfied by any theoretical
galaxy formation model. 

A generic outcome of theoretical models such as these is that the amount of cold 
gas that can be transformed to stars increases as a function of the mass slower
than the halo mass itself, because larger halos are hotter and the
gas takes longer to cool \cite{bias,Kauffmann,Benson}. 
In such models one would expect 
$\langle N \rangle/M$ to decline with $M$. This is shown in 
figure \ref{fig3} where $\langle N \rangle$ and $\langle N(N-1)\rangle^{1/2}$
is plotted versus $M$ for galaxies selected only on the basis on 
absolute magnitude ($M_B<-19.5$). Both functions have similar dependence
for $M>10^{14}h^{-1}M_{\sun}$.
When the number of galaxies per halo begins to drop below unity
the two functions begin to deviate from one another and
$\langle N(N-1)\rangle^{1/2}$ drops below $\langle N \rangle$. This is because
only the halos with two or more galaxies contribute to 
$\langle N(N-1)\rangle^{1/2}$, while single galaxy halos also 
contribute to $\langle N \rangle$. However, both functions 
increase less rapidly than the mass for $M>10^{13}h^{-1}M_{\sun}$.

\begin{figure}
\centerline{\psfig{file=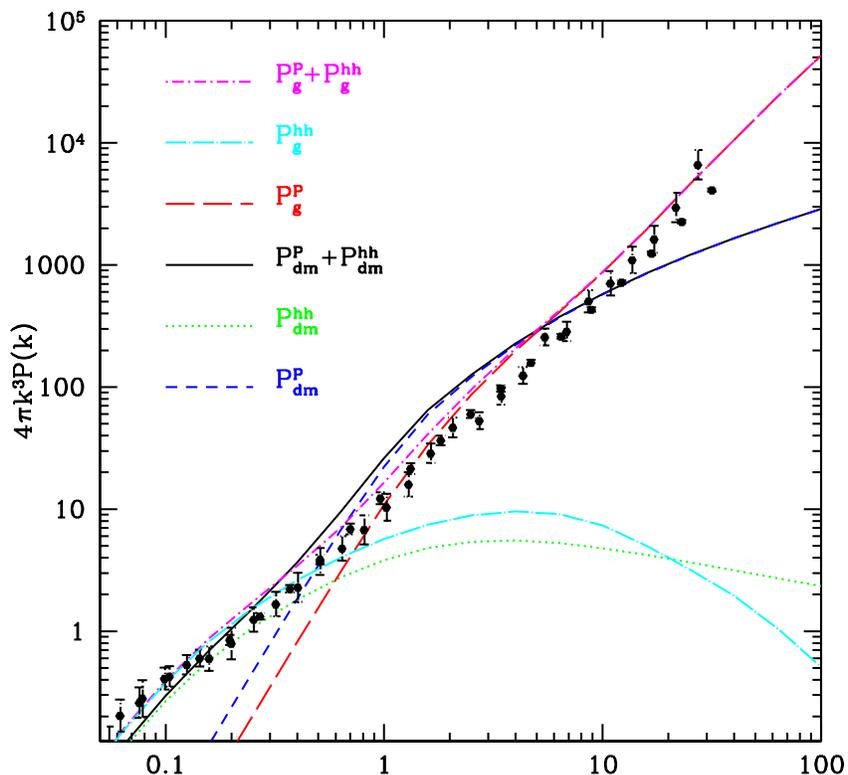,width=4.5in}}
\caption{Comparison between galaxy and dark matter power spectrum 
predictions for galaxies selected by absolute magnitude $M_B<-19.5$
as in figure \ref{fig3}. Poisson, halo-halo and 
combined terms are shown for the two spectra.
Also shown is the measured power spectrum of galaxies.
Note that at low $k$ the Poisson term for the galaxies is lower than that 
for the dark matter and this delays the onset of nonlinear clustering 
in galaxies.
}
\label{fig4}
\end{figure}

Using $\langle N(N-1)\rangle^{1/2}/M $ and $\langle N\rangle /M$ for 
$M_B<-19.5$ from 
figure \ref{fig3} in equations 
\ref{ngal}-\ref{gp} we obtain the galaxy power spectrum 
shown in figure \ref{fig4}. We only show results for $\alpha=-1$
model, but the $\alpha=-1.5$ model gives essentially identical results.
Also shown is the dark matter power spectrum and its two contributions, 
as well as the measured APM and scaled IRAS galaxy power spectrum compiled in \cite{P97}. 
First thing to note is the good agreement between our analytical 
model and the simulations. The agreement is  
significantly better for this model than for the model where there
is no central galaxy, which would give a stronger decline in power 
on small scales.
The galaxy power spectrum is 
almost a perfect power law over several decades in scale, in agreement 
with observations and in contrast to the dark matter power spectrum, 
whose slope gradually decreases with $k$. 
The slope of the galaxy power spectrum is in agreement with the observed slope 
$k^3P(k)\propto k^{1.8}$ and this slope persists in the analytic model 
down to kpc scales. 

It is useful to introduce bias $b(k)$,  defined as the square root of the 
ratio between galaxy and dark matter power spectrum,
\begin{equation}
b(k)=[P_{\rm gg}(k)/P_{\rm dm,dm}(k)]^{1/2}. 
\end{equation}
The bias $b(k)$ is approximately constant and close to unity 
on large scales, decreases and becomes less than 
unity between $0.3{\rm hMpc}^{-1}<k<6{\rm hMpc}^{-1}$ and then increases 
for large $k$. The bias
is therefore scale dependent and nonmonotonic, both of which as shown below 
are generic predictions of this model. 
On very large scales the power spectrum is dominated by the correlations between 
the halos and the internal structure of halos can be neglected. This gives
the constant bias on large scales, which for the galaxy type considered 
here is close to unity. On smaller scales the halo Poisson term 
becomes important both for galaxies and dark matter. However, if 
$\langle N(N-1)\rangle^{1/2}/M  \propto M^{\psi}$ with $\psi<0$
the Poisson term for galaxies is lower than the Poisson term for dark matter
in the limit $y(k,M)=1$. 
This is because the halo Poisson term is larger 
if halos are rarer.
If $\psi<0$ the dominant contribution in galaxy 
power spectrum is shifted to lower mass halos, which are more abundant and
this reduces the Poisson term relative to dark matter. 
Another important factor that reduces the galaxy Poisson term is that $\langle N \rangle$
exceeds $\langle N(N-1)\rangle^{1/2}$ below $M \sim 10^{14}h^{-1}M_{\sun}$. 
$\langle N \rangle(M)$ determines the mean density of galaxies $\bar{n}$ in 
equation \ref{ngal}. This suppresses the Poisson term in equation \ref{gp}
even if $\psi=0$. Suppression of the galaxy Poisson term relative to the dark matter 
delays the onset of nonlinear power in the galaxy power spectrum relative to 
the dark matter, which is clearly seen in figure \ref{fig4}.
It gives a natural explanation for the position of the inflection point 
in the observed galaxy power spectrum
without the need to introduce phenomenological double power law spectra
\cite{P97}. While our model is already in a good agreement with the data 
an even better fit would be achieved with a somewhat smaller Poisson term, 
which would require $\psi$ to be even lower or $\langle N \rangle$
to exceed $\langle N(N-1)\rangle^{1/2}$ even more. This would further delay the onset 
of the nonlinear clustering. 

\begin{figure}
\centerline{\psfig{file=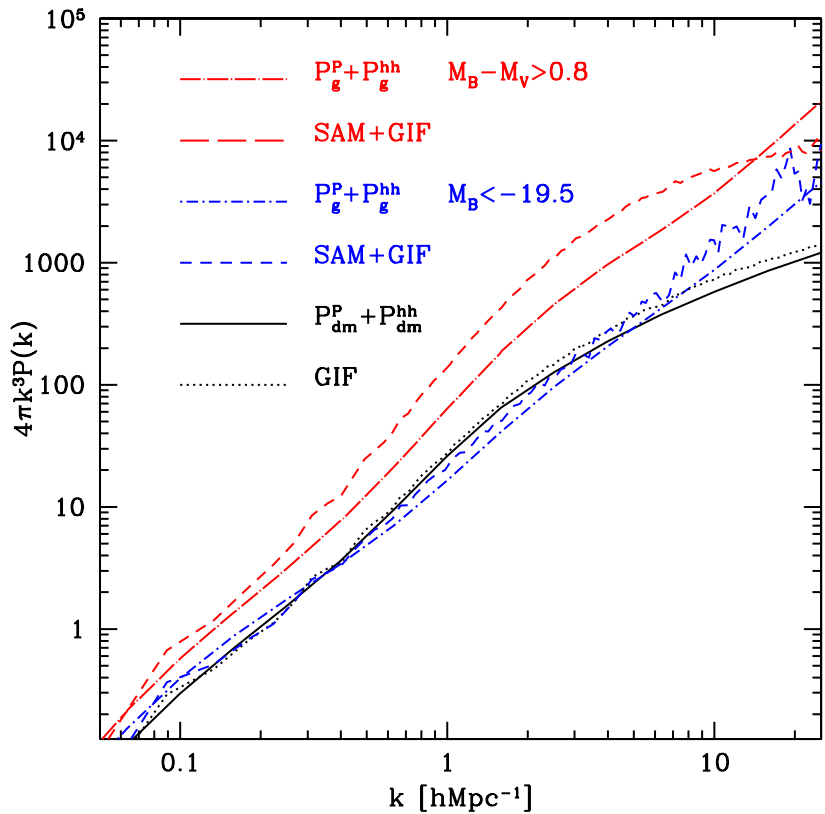,width=4.5in}}
\caption{Comparison between galaxy and dark matter power spectrum 
predicted by our model and the results of
N-body simulations and semi-analytic models.
Predictions for galaxies selected by absolute magnitude $M_B<-19.5$
and $M_B-M_V>0.8$ are shown.
}
\label{fig4b}
\end{figure}

On even smaller scales the halo profile $y(k,M)$
becomes important, since it begins to decrease from unity 
at a scale that corresponds to a typical size of the halo, which is 
smaller for the lower mass halos. Since the galaxy power spectrum 
is weighted towards lower mass halos relative to dark matter the 
term $y(k,M)$ begins to be important in suppressing the Poisson term 
at a smaller scale. In addition, if each halo hosts a central galaxy 
then switching to $p=1$ for small halos also makes the suppression
by the halo profile less important. 
Smaller suppression of
the galaxy power spectrum relative to the dark matter results in an increase
of bias with $k$. 
This argues that the decrease of $b(k)$ on intermediate scales and 
the increase on small scales are generic predictions. The 
overall result of this is an approximate power law
of the galaxy power spectrum over several decades. Such a power law 
arises quite generically in a CDM family of models where $\psi<0$
(or $\langle N \rangle> \langle N(N-1)\rangle^{1/2}$)
and where each halo hosts a central galaxy. We note that
the latter is required 
to preserve the power law behaviour to very small scales. A model where 
$p=2$ for all halo masses turns below the power law 
in the power spectrum at $k>50h$Mpc$^{-1}$, 
similar to the dark matter.

The conclusion above that the bias first declines with $k$ and then 
rises again applies to
a normal galaxy population.
If one selects red galaxies on the basis of
color or ellipticals on the basis of morphology then one may expect a
different bias dependence, since red or elliptical galaxies are 
preferentially found in more massive halos, such as groups and clusters.
Figure \ref{fig3} shows that the galaxies 
selected by $M_B-M_V>0.8$ and $M_B<-19.5$ are dominant in halos 
with $M > 10^{14}h^{-1}M_{\sun}$, while their relative fraction declines
rapidly below that. 
The dependence of $\langle N(N-1)\rangle^{1/2} $ with $M$ 
is much steeper in this case so that 
$\psi >0$ for $M < 10^{14}h^{-1}M_{\sun}$. In addition  
$\langle N \rangle \sim  \langle N(N-1)\rangle^{1/2}$
across the entire range of halo masses, a consequence of the fact that 
most of the red galaxies are not central galaxies, which in these
models show recent star formation and are therefore not red. 

Figure \ref{fig4b} shows the comparison between analytic predictions 
and results from semi-analytic models \cite{Kauffmann}. We use
mass dependence from figure \ref{fig3} in equations 
\ref{ngal}-\ref{gp} for both $M_B<-19.5$ and $M_B-M_V>0.8$ 
galaxy selection.
Also shown are the dark matter power spectrum from the model and 
from the GIF simulations \cite{Jenkins} which were used for 
semi-analytic models.
Qualitatively the agreement is excellent,
specially for dark matter and $M_B<-19.5$ galaxies,
while for red galaxies semi-analytic models predict a somewhat higher amplitude.
Part of the disagreement is caused simply 
by dark matter spectrum not being in agreement with PD (our models are 
chosen so that they agree with PD) \cite{Jenkins}. We do not show small scales
($k>20h$Mpc$^{-1}$) where
limited resolution of N-body simulations prevents a meaningful comparison.
The remaining discrepancy for red galaxies between $0.5h$Mpc$^{-1}<k<20h$Mpc$^{-1}$
can only be explained by them not tracing exactly dark matter 
distribution in halos with $M>10^{13}h^{-1}M_{\sun}$. The red
galaxies must be more centrally concentrated than dark matter 
in semi-analytic models in order that their power spectrum has a 
higher amplitude than predicted from our model. This is in 
agreement with direct analysis of galaxy distribution inside halos
using the same simulations \cite{Diaferio99}, where it was found
that red galaxies in $\Lambda CDM$ model tend to be more centrally 
concentrated that dark matter. Galaxies that form first end up more 
towards the center of the cluster because the violent relaxation during the 
merging is incomplete.

In the case of the red galaxies the bias starts with a value larger 
than unity on large scales. This is because most of the red galaxies 
are in clusters which are biased relative to the dark matter following 
equation \ref{b}. Bias first rises with $k$ and then 
declines. This is just the opposite from the scale dependence of the
normal galaxies and is a consequence of $\psi>0$
for $M < 10^{14}h^{-1}M_{\sun}$ and
$\langle N \rangle \sim	 \langle N(N-1)\rangle^{1/2}$. This
gives rise to the Poisson 
term larger for the galaxies than for the dark matter on large scales. 
This conclusion is again independent of the 
distribution of the galaxies inside the halos. 
This is confirmed in figure \ref{fig4b} where on large scales 
our model
agrees very well with the semi-analytic predictions.
Because the galaxies are preferentially 
in larger halos relative to the dark matter $y(k,M)$ suppression is 
more important and the bias 
declines on smaller scales. This is seen in the power spectrum
from the simulations. In our model it begins to rise again 
on even smaller scales because  
$p$ switches to unity for $M< 10^{14}h^{-1}M_{\sun}$, 
resulting in a smaller suppression by the halo profile.
This effect is not seen in the simulations, presumably because of 
their limited resolution. 

It is important to note that  
bias may never be really constant even 
on scales above $100h^{-1}$Mpc. For the red sample
it changes 
by 30\% between $k=0.01h$Mpc$^{-1}$ and $k=0.1h$Mpc$^{-1}$. 
This is because the Poisson term 
does not become much smaller than the halo-halo term even on very 
large scales, a consequence of the fact that the slope of 
$P_{\rm lin}(k)$ and 
thus the halo-halo term itself becomes flat and even positive on very 
large scales  
(approaching $n \sim 1$ on very large scales). 
Since even at the turnover of the power 
spectrum (where $n \sim 0$) the Poisson term 
for the red galaxies is of the order of 20\% of the halo-halo term the bias
does not become constant and begins to increase again on scales 
larger than the scale of the turnover. In fact on very large scales 
($k<10^{-3}h$Mpc$^{-1}$) the 
red galaxy power spectrum becomes white noise, although these 
scales are already approaching the size of the visible universe. It should be 
noted that this description is valid on large scales only
for galaxies which do not 
obey mass and momentum conservation. For the dark matter mass and 
momentum conservation require that the Poisson term vanishes on 
large scales and any spectrum generated by a local process should 
decrease faster than $P(k) \propto k^4$ as $k \rightarrow 0$
\cite{zeldovich}. Galaxies do not obey mass and momentum 
conservation and can have the Poisson contribution, so 
the qualitative scale dependence of bias remains as predicted above.

We have concentrated on the power spectrum above because it is 
the quantity that can be most directly compared to the theoretical
predictions. The same analysis could however be applied to the 
correlation function as well. The power law dependence of the 
power spectrum would also result in a power law correlation function, 
so the conclusions would remain unchanged. The main difference in the real 
space is that the Poisson term is localized to scales smaller than 
the typical halo scale and vanishes on scales above that. In this 
case bias would be scale dependent up to this typical scale 
(of order few Mpc),
but would become scale independent on scales above that. 
There is no need to model the Poisson term on large scales at all. 
In this sense the real space correlation function offers
some advantages over the power spectrum, where one must
attempt to remove the Poisson term in the power spectrum by 
modelling it as a constant term on large scales.

Our predictions agree with the results 
of semi-analytic models, indicating that the here proposed model  
is sufficient to extract the key ingredients to model 
the galaxy clustering. 
This means one does not need to rely on N-body simulations as 
long as the ingredients of the model are specified.
If one can extract $\langle N \rangle$, 
$\langle N(N-1)\rangle^{1/2}(M)$ and $y(k,M)$ directly from the data 
one can sidestep the theoretical modelling of this relation and 
predict the galaxy power spectrum directly \cite{KNH}. 
It is in principle possible 
to obtain such information at least for the massive halos 
by combining dynamical information 
on galaxy groups and clusters, such as X-ray temperature, 
velocity dispersion or weak lensing mass, with the number 
of galaxies in these clusters. Existing data such as 
CNOC survey \cite{Carlberg} indeed find that $\langle N \rangle/M$ for 
galaxies with $M_K<-18.5$ is systematically 
lower in massive clusters with $\sigma>1000$ km/s than in 
poorer clusters. The current data are sparse, but 
new large surveys such as SDSS
and 2dF will enable one to extract such information with a much 
better statistics. This could allow one to determine within our model the
dark matter power spectrum from the galaxy power spectrum directly.

Another direction to obtain $\langle N \rangle/M$ is to require 
consistency with other measurements that combine dynamical 
and galaxy information. Galaxy-dark matter correlations 
discussed in the next section are one possibility. Another are
pairwise velocity dispersion measurements. If
$\langle N \rangle/M$ declines with $M$ then the pairwise velocity
dispersion for the galaxies will be lower than for the dark matter 
\cite{JingMoBo98,Benson99b}. This 
is because there will be more pairs of galaxies in smaller halos 
relative to the dark matter. Smaller halos have
smaller velocity dispersions and smaller relative velocities between 
the particles.
This can explain the lower amplitude of pairwise velocity dispersion 
in the LCRS data compared to the N-body simulations \cite{JingMoBo98}. The 
required value of $\psi \sim -0.1$ has indeed the same sign as required
to reproduce the delayed onset of nonlinear clustering and the power 
law in galaxy power spectrum. 
It would be interesting to see whether a single set of 
functions  $\langle N \rangle (M)$,  $\langle N (N-1)\rangle^{1/2}(M)$
can provide a unified 
description to both galaxy clustering and pairwise velocities within
the CDM models.

\section{Dark matter-galaxy cross-correlation}
Dark matter-galaxy cross-correlations are measured whenever 
a galaxy is cross-correlated with a tracer of the dark matter. 
Examples of this are 
galaxy-galaxy lensing \cite{galgal}, where one is measuring correlation
between galaxies and cosmic shear, and correlations between foreground and
background galaxies or quasars \cite{qsogal}, 
where correlations (or anti-correlations) are 
induced by magnification bias of background 
objects. In both cases one is measuring the correlations between the
galaxies and dark matter along the line of sight, which can be expressed 
as a convolution over the galaxy-dark matter cross-correlation power spectrum.

Galaxy-dark matter cross-correlations have been modelled in the past
using either a bias model relating them to the dark matter or galaxy 
power spectrum \cite{waerbeke} or using galaxies 
sitting at the centers of the galactic size halos \cite{brainerd}. 
In the first description assuming galaxy-dark matter cross-correlations
measure bias $b(k)P_{\rm dm}(k)$, which in combination with the
galaxy power spectrum $b^2(k)P_{\rm dm}(k)$ can give both $b(k)$ and $P_{\rm dm}(k)$.
Such a model is a reasonable description on large scales, but must 
break down on small scales where galaxies do not trace dark matter and 
there is no guarantee that the scale dependent bias that relates 
$P_{\rm gal,dm}(k)$ and $P_{\rm gal}(k)$ can be used to extract
$P_{\rm  dm}(k)$.

Second model describes cross-correlations in terms of 
galaxies sitting at the centers of their halos and interprets the 
results in terms of the averaged halo profile \cite{brainerd}. 
There are two potential problems with this approach. First, 
there may be more than one galaxy inside the halo, which is 
specially important for large halos (figure \ref{fig3}). 
Since not all galaxies can lie at the halo center this can affect the 
interpretation of the cross-correlations in terms of the halo profile.
Second, just as in the case of the dark matter  
the contribution to the power spectrum
comes from a range of halo masses and one cannot model 
the galaxy-dark matter cross-correlation
simply as a typical $L_*$ galaxy halo profile. The 
strength of the correlations is determined both by the dark 
matter profile of the halos as well as by the halo mass function, 
so the slope of the correlation function that one is ultimately 
measuring with galaxy-galaxy lensing and foreground-background galaxy 
correlations need not be directly related to the dark matter profile 
\cite{Guzik00}.
Model developed in previous sections may be applied to the 
dark matter-galaxy cross-correlation
power spectrum to quantify these issues in more detail. 

Galaxy-dark matter cross-correlation power spectrum 
has halo-halo and halo Poisson terms. 
First term describes the correlations between galaxies and 
dark matter in neighbouring halos and is dominant on 
large scales. Second term includes the correlations between the galaxies
and dark matter in the same halo and dominates on small scales.
The halo-halo term is given by 
\begin{equation}
P^{hh}_{\rm g,dm}(k)=P_{\rm lin}(k) \left[{\bar{\rho}\over \bar{n}}
\int f(\nu)d\nu  {\langle N \rangle \over M}b(\nu)y[k,M(\nu)]\right] 
\left[\int f(\nu)d\nu  b(\nu)y[k,M(\nu)]\right]. 
\label{chh}
\end{equation}
On large scales where this term dominates it reduces to constant 
bias model, $P^{hh}_{\rm g,dm}(k)=\langle b \rangle P_{\rm lin}(k)$. 
The Poisson term in the model where galaxies trace dark matter 
inside the halos except for the central galaxy sitting at its center 
is given by 
\begin{equation}
P^P_{\rm g,dm}(k)= {1 \over (2\pi)^3 \bar{n}}
\int f(\nu)d\nu \langle N \rangle 
|y(k,M)|^p.
\label{cp}
\end{equation}
Here $p=2$ for $\langle N \rangle >1$ and $p=1$ 
for $\langle N \rangle <1$.

\begin{figure}
\centerline{\psfig{file=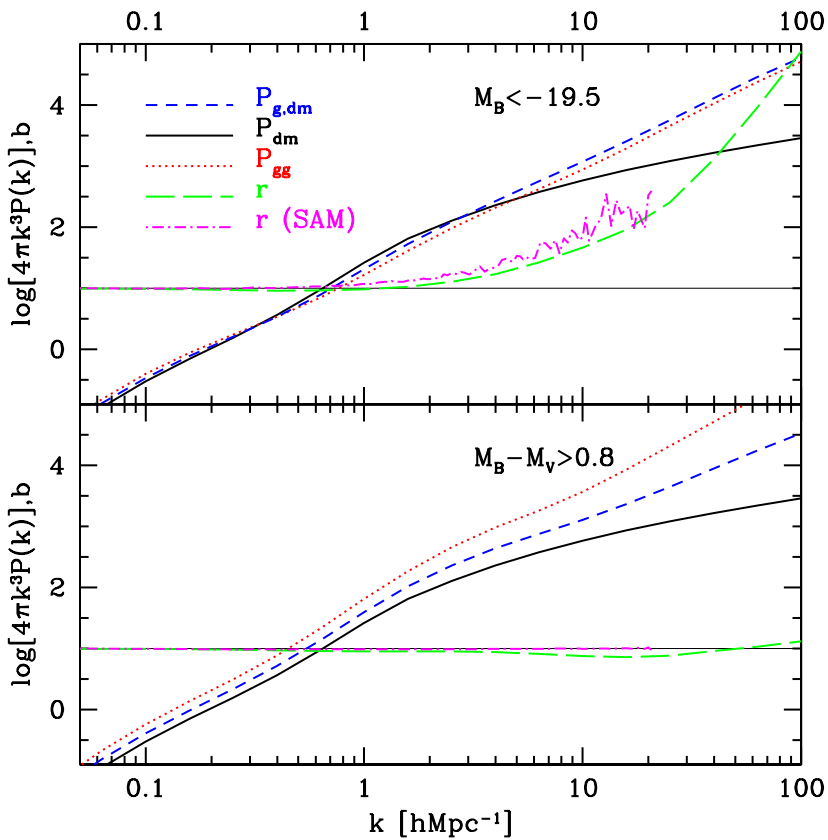,width=4.5in}}
\caption{Galaxy-dark matter cross-correlation power 
spectrum for the two galaxy types as in figure \ref{fig3} (dashed),
together with the dark matter (solid) and galaxy (dotted) power 
spectrum. Also shown is the cross-correlation coefficient $r(k)$
from the model (dash-dotted) and from simulations (long-dashed).
}
\label{fig5}
\end{figure}

Figure \ref{fig5} shows the results for the cross-correlation power 
spectrum for the same galaxy selection as in figures \ref{fig3}
and \ref{fig4}. For regular galaxies selected by an absolute 
magnitude (top panel) the cross-correlation spectrum is 
similar to the galaxy power spectrum. 
If we define the cross-correlation coefficient as
\begin{equation}
r(k)={P_{\rm g, dm}(k) \over [P_{\rm dm}(k) P_{\rm g}(k)]^{1/2}},
\end{equation}
then we see from figure \ref{fig5} that it is approximately 
unity up to $k \sim 1h$Mpc$^{-1}$ and increases for higher $k$.
Note that the cross-correlation coefficient is not restricted 
to $|r(k)|<1$ because we have subtracted out the shot noise term 
from the galaxy power spectrum following the usual approach \cite{Peebles}.
Because on small scales the galaxy and cross-correlation spectra 
are comparable and exceed the dark matter spectrum the cross-correlation
coefficient 
grows to large values in this model. Comparison with the semi-analytic results 
\cite{Guzik00} again shows very good agreement up to the resolution 
limit of the simulations. 
Bottom of figure \ref{fig5} shows the results for the red galaxies.
In this case the cross-correlation spectrum falls in between the
dark matter and the galaxy spectrum, so that $r \sim 1$ down to 
very small scales. This is again in agreement with semi-analytic
results which show $r\sim 1$ throughout the entire range of $k$.

The main reason for $r \ne 1$ on small scales is that
$\langle N(N-1)\rangle^{1/2} \ne \langle N\rangle$ (figure 
\ref{fig3}). The 
difference between the two functions is 
more significant  
for the normal than for the red galaxies,
which is why the 
cross-correlation coefficient begins to deviate from unity at larger 
scales for $M_B<-19.5$ than for $M_B-M_V>0.8$. 
Because in this regime
$\langle N(N-1)\rangle^{1/2} <\langle N\rangle$ this leads to 
$r(k)>1$, as seen in figure \ref{fig5}. It is interesting to note
from figure \ref{fig3} that for the red galaxies the two functions
agree very well even below unity and this leads to $r(k) \sim 1$ down to very 
small scales. When this happens one can reconstruct the dark matter 
power spectrum from the galaxy and cross-correlation spectrum even if 
most of the dark matter halos are not directly observed. 
Unfortunately one cannot extract these two functions without 
first identifying the dark matter halos, so this prediction
cannot be directly verified from 
observational data using the galaxy information only.

Second source of stochasticity is the presence of central galaxy.
For those halos where 
$\langle N(N-1)\rangle^{1/2}<1$ or 
$\langle N\rangle <1$ only one power of $y(k,M)$ is used 
as opposed to two in the case of the dark matter. This 
induces some stochasticity even if $\langle N\rangle =
\langle N(N-1)\rangle^{1/2} $, because it enhances the galaxy-galaxy and 
galaxy-dark matter spectrum above the dark matter-dark matter spectrum. 
Another source of stochasticity
would be $\psi \ne 0$, which would make correlations at a given scale
being dominated by different mass range 
in the case of the dark matter and the galaxies. 
Calculations where only this effect is present 
give $r \sim 1$
over a wide range of scales, showing that this cannot 
be a significant source of stochasticity, at least for reasonable 
values of $\psi$.

\begin{figure}
\centerline{\psfig{file=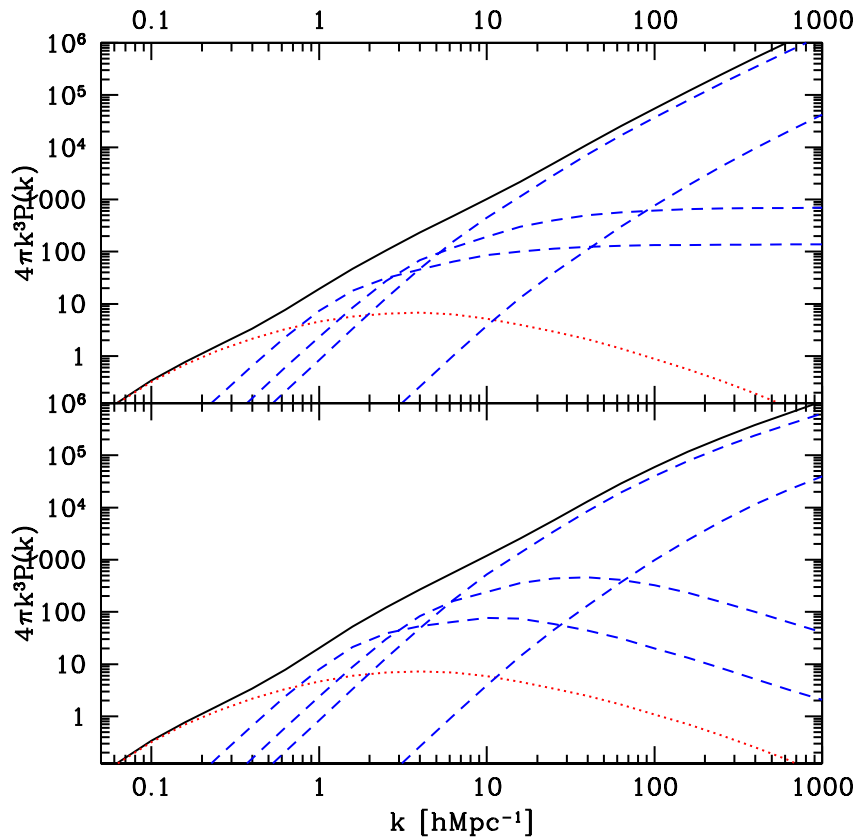,width=4.5in}}
\caption{Contribution to the dark matter
galaxy cross-correlation power spectrum from the different 
halo mass intervals. The curves correspond to the same mass
intervals 
as in figure \ref{fig2}.
Top is $\alpha=-1.5$ model, bottom $\alpha=-1$. 
}
\label{fig6}
\end{figure}

Our model predicts that even if the constant bias model is not valid, 
its generalization $r=1$ model may be
a reasonable approximation at least down to 1 Mpc scales.
An example are the red galaxies (bottom of figure \ref{fig5}), which have very 
strong scale dependent bias, yet $r\approx 1$ over a wide range 
of scales.  In this sense 
determining the dark matter power spectrum from the measurements of galaxy-galaxy 
spectrum and galaxy-dark matter spectrum 
under the assumption of $r=1$ may have a larger range of validity 
than the constant bias model. This relies on the assumption 
$\langle N\rangle=\langle N(N-1)\rangle^{1/2}$ predicted 
from these models. This prediction can be 
verified at least for the more massive halos directly from observations, 
for example by using galaxy counts in cluster catalogs to extract 
$\langle N\rangle$ and $\langle N(N-1)\rangle^{1/2}$.
Such an approach would provide an alternative way to determine $r(k)$ 
directly from the data.

The model developed here can also be used to clarify the
interpretation of the galaxy-dark matter cross-correlation in 
terms of an averaged density profile of a typical galaxy.
Figure \ref{fig6} shows the contribution to the 
cross-correlation spectrum from the different halo mass intervals, 
similar to figure \ref{fig2} for the dark matter. For
$k<20h$Mpc$^{-1}$, corresponding approximately to scales larger 
than 100h$^{-1}$kpc in real space,
one cannot interpret the correlations in terms of the
shape of a single halo profile, but instead  
as the convolution of these over the halo mass function, multiplied
with the number of galaxies per halo. 
Observed correlations on large scales do not necessarily mean 
that the halo of an $L_*$ galaxy extends to large distances. 
Instead, it is more likely that one is observing correlations 
arising from the group and cluster size halos, which exceed the 
correlations contributed
from the galactic size halos on larger scales.
This cannot be corrected in any simple manner 
by taking into account 
the correlation function of the galaxies \cite{galgal}, which attempts to
model the 
presence of other nearby halos. Even if the galaxy correlations
vanished  one would still need to take into account the halo mass 
function and the fact that different halos dominate on different 
scales. 
More detailed discussion of these
points will be presented elsewhere \cite{Guzik00}.

On smaller 
scales the transition to $\langle N\rangle(M) <1$ implies that 
$y(k,M)$ suppression is less important because  
$p=1$.
This is further enhanced by the flattening of $\langle N\rangle$
below $M\sim 10^{13}h^{-1}M_{\sun}$ as seen in figure \ref{fig3}.
In addition, galaxies selected on the basis of their absolute magnitude
cannot exist in very small halos, so the mass function has a 
strong cutoff below $10^{12}h^{-1}M_{\sun}$. 
Thus on scales with $k>20h$Mpc$^{-1}$
the galaxy-dark matter cross-correlation may be better interpreted
in terms of the average profile of 
$10^{12}h^{-1}M_{\sun}<M< 10^{13}h^{-1}M_{\sun}$ 
halos.
However, this may not be a robust prediction since
the semi-analytic predictions in figure \ref{fig3}
are highly uncertain over this mass range. A small change
in $\langle N \rangle (M)$ may lead to a larger influence of
the mass function on the power spectrum,
making the correspondence between the halo profile and the power spectrum
less certain. In general one should be cautious in interpreting 
the shape of the
galaxy-dark matter correlation function in terms of an averaged 
dark matter profile. 

\section{Conclusions}
We developed an analytic model for computing 
the power spectrum of the dark matter, galaxies
and their cross-correlation based on the Press-Schechter model. In 
this model 
all the matter in the universe is divided into virialized halos.
These halos cluster and have some internal profile. The total 
power spectrum is the sum of the halo clustering term and the halo 
Poisson term, which accounts for the correlations within the halos. 
We assume that the halo profiles are self-similar regardless of the
initial conditions, but with the
mass dependent concentration parameter, as suggested by
high resolution simulations \cite{NFW,Kravtsov,Moore}. 
The model agrees well with the
results of N-body simulations for the $\Lambda CDM$ model.
We are able to find a good agreement for inner slopes $\alpha=-1$
and $\alpha=-1.5$, indicating that the shape of the nonlinear power 
spectrum cannot by itself distinguish between the two. 

The model can in principle be applied to any cosmological model, including 
those with a cutoff in the linear power spectrum on small scales 
or with some features in the power spectrum.
While this will be explored in more detail in a future paper we wish to 
emphasize here that the mass function, which is sensitive to the 
linear power spectrum, has a direct effect on the nonlinear power 
spectrum through the halo abundance, 
so that not all of the information on the linear power 
is lost in the nonlinear regime. For example, if the linear power 
spectrum is cut-off on small scales and if inner profile $\alpha>-1.5$  
as suggested by the simulations then the correlation function or $k^3P(k)$ 
must have a turnover on small scales. 
This differs from the CDM models which predict the nonlinear correlation 
function to continue to grow on small scales. If
we wish to eliminate the halos with $M<10^{11}h^{-1}M_{\sun}$ \cite{LK} then this
would suppress the power on scales below 10kpc (figure \ref{fig2}). 
This effect therefore becomes significant on scales smaller 
than those resolved in a recent study of such truncated power spectrum 
models \cite{White}.

Our main conclusion regarding the galaxy power spectrum is that a 
simple model for the dependence of the linear and pair weighted number of 
galaxies inside halo as a function of the halo mass can explain most of
the properties of the galaxy clustering seen in more complicated 
models based on the N-body simulations. A power law in the galaxy 
correlation function with slope $1.8$ is a generic prediction
of the model where the number of galaxies inside the halo increases 
less rapidly with mass than the halo mass itself, mean number of 
galaxies exceeds pair weighted average and there is a central galaxy
in each halo. 
The decline of number of galaxies per unit mass as a function of mass
is predicted by 
the galaxy formation models \cite{Benson,Kauffmann,bias} 
and has been observed in clusters \cite{Carlberg}.
It is also required to explain the pairwise velocity dispersion results 
\cite{JingMoBo98}.
For such galaxies bias
first decreases below unity, because the 
Poisson term is smaller for them than for dark matter. 
This naturally explains the later onset of nonlinearity in 
galaxy power spectrum compared to the dark matter, which 
reconciles the discrepancy between the data and 
the CDM models \cite{P97}.
Conversely, there is no need to invoke poorly motivated models
such as double power law model \cite{P97}.
On large scales bias converges to a constant for these galaxies.

Red or elliptical galaxies, which 
are more abundant in massive halos, show a different relation: 
their number inside the halos increases on average 
more rapidly than the halo mass. 
In this case bias increases with $k$ above the turnover in the power 
spectrum ($k \sim 0.01h$Mpc$^{-1}$), because
their Poisson term is larger than 
that of dark matter. In fact, the Poisson term may be so strong that 
it may not be negligible compared to the halo clustering term 
even on very large scales and one may not converge to
the constant bias model.

Galaxy-dark matter correlations can also be predicted by this 
model. In this case one must specify 
the average number of galaxies per halo 
as a function of halo mass. 
Here again our model reproduces the main features present in the 
N-body simulations with semi-analytic galaxy formation \cite{Guzik00}. 
Galaxy-dark matter cross-correlations can be measured with 
galaxy-galaxy lensing or correlations between foreground and background 
galaxies and may provide a way to break some 
of the uncertainties present with the galaxy clustering. 
For example, we have shown that even if the constant bias may not be
a good approximation, cross-correlation coefficient may nevertheless
be close to 
unity down to Mpc scales, which 
would allow one to extract the dark matter power spectrum 
from the knowledge of the galaxy and cross-corelation spectrum 
on scales larger than this. The main source of stochasticity ($r \ne 1$)
arises from the pair weighted number of galaxies inside the halo 
differing from the mean number of galaxies and from the (possible) 
existence of central galaxies in the halos. 

We have emphasized that caution must be applied when 
interpreting the cross-correlations such as galaxy-galaxy 
lensing in terms of an averaged
density profile of a halo. As we have shown different halo 
masses dominate on different scales and the correlation 
function reflects this combined effect of all the halos. 
For example, correlations at 
a few hundred kpc observed by galaxy-galaxy lensing \cite{galgal} 
are more likely to be caused by group and cluster 
sized halos at $r_s$ distances than by galaxy sized halos at 
$r_v$ distances. More detailed work is needed to extract
the structure and extent of the dark matter halos from such observations.

Perhaps the most promising direction to explore in the future 
is to extract the functional dependences that parametrize our 
model directly from the observations.
If one can determine the linear and pair weighted number of galaxies as a function 
of halo mass and their distribution inside the halos then one can
determine the galaxy power spectrum directly within this model. 
Similarly if one can compare the mean number of galaxies with the
pair weighted number as a function of halo mass then one can 
predict the galaxy-dark matter cross-correlation coefficient.
This is certainly  feasible for clusters, which dominate the Poisson term 
on large scales. Current data are sparse \cite{Carlberg}, but
new surveys such as SDSS or 2dF should provide sufficient 
statistics to make this feasible. This approach would provide an independent
estimate of the scale dependence of bias and correlation coefficient 
on large scales. It will also provide important constraints that would
need to be satisfied by any viable galaxy formation model.

I ackowledge the support of NASA grant NAG5-8084. 
I thank G. Kauffmann and S. White for a detailed reading of the manuscript
and for providing results of GIF
N-body and
semi-analytic simulations 
and J. Guzik for help with them. I also thank
R. Sheth and R. Scoccimarro for useful conversations and 
help in initial stages of this project and
J. Peebles and U. Pen for useful discussions.


\begin{thebibliography}{99}

\bibitem{DS} A. Dekel and J. Silk, \apj {\bf 303}, 39 (1986).
\bibitem{SS99}
D. N. Spergel and P. J. Steinhardt, preprint astro-ph/9909386 (1999).
\bibitem{P97} see e.g. a compilation in 
J. A. Peacock, \mnras{\bf 284} 885 (1997).
\bibitem{qsogal} a review is given in 
M. Bartelmann and P. Schneider, submitted to Phys. Rep.,
preprint astro-ph/9912508 (1999).
\bibitem{brainerd}
T. G. Brainerd, R. D. Blandford and I. Smail, \apj {\bf 466} 623 (1996).
\bibitem{PS74}
W. H. Press and P. Schechter,
Astrophys. J. {\bf 187}, 425 (1974).
\bibitem{MS} J. McClelland and J. Silk, \apj,
{\bf 217}, 331 (1977).
\bibitem{SJ} R. Sheth and B. Jain, \mnras {\bf 285}, 231 (1997).
\bibitem{NFW}
J. Navarro, C. Frenk and S. D. M. White, \apj {\bf 462}, 563 (1996).
\bibitem{ScoSh} Similar approach has been developed independently by 
R. Scoccimaro and R. Sheth (in preparation, 2000).
\bibitem{bias} M. Blanton, R. Cen, J. P. Ostriker and M. A. Strauss 
\apj {\bf 522} 590 (1999); F. R. Pearce et al., \apj {\bf 521} L99 (1999). 
\bibitem{Benson} A. J. Benson, S. Cole, C. S. Frenk, C. M. Baugh and 
C. G. Lacey, \mnras in press, astro-ph/9903343 (1999).
\bibitem{Kauffmann} G. Kauffmann, J. M. Colberg, A. Diaferio and S. D. M. 
White, \mnras {\bf 303} 529 (1999); {\it ibid} \mnras {\bf 307} 529 (1999).
\bibitem{biasobs} see e.g. a compilation in
J. A. Peacock and S. J. Dodds, \mnras{\bf 267} 1020 (1994).
\bibitem{JingMoBo98} Y.P. Jing, H.J. Mo and G. Boerner, \apj {\bf 494},
1 (1998).
\bibitem{Benson99b} A. J. Benson, C. M. Baugh, 
S. Cole, C. S. Frenk and C. G. Lacey,
\mnras submitted, astro-ph/9910488 (1999).
\bibitem{galgal} I. P. Dell'Antonio and J. A. Tyson, \apj {bf 473}, L17 (1996); 
M. J. Hudson, S. D. J. Dwyn, H. Dahle and N. Kaiser, 
\apj {\bf 503} 531 (1998); R. E. Griffiths, S. Casertano, M. Im and 
K. U. Ratnatunga, \mnras {\bf 282} 1159 (1996); Fischer et al. astro-ph/9912119
(1999).
\bibitem{waerbeke}
L. van Waerbeke, \aa {\bf 334} L1 (1998).
\bibitem{Huss} A. Huss, B. Jain and M. Steinmetz, \apj {\bf 517} 64 (1999).
\bibitem{Tormen} G. Tormen, F. R. Bouchet and S. D. M. White, \mnras {\bf 286}
865 (1997).
\bibitem{Moore} B. Moore, , F. Governato, T. Quinn, J. Stadel and G. Lake,
\apj {\bf 499} L5 (1998);
B. Moore, T. Quinn, F. Governato, J. Stadel and G. Lake,
astro-ph/9903164 (1999). 
\bibitem{Fukushige} T. Fukushige and J. Makino, \apj {\bf 477} L9 (1997).
\bibitem{Kravtsov} A. V. Kravtsov et al., \apj {\bf 502} 48 (1998).
\bibitem{JingSuto99} Y. P. Jing and Y. Suto, \apj in press, astro-ph/9909478
(1999).
\bibitem{Bullock99} J. S. Bullock, \mnras submitted, astro-ph/9908159 (1999).
\bibitem{Somm} M. A. K. Gross et al., \mnras {\bf 301}, 81 (1998);
R. S. Somerville, G. Lemson, T. S. Kolatt and A. Dekel, \mnras, in press
(astro-ph/9807277) (2000).
\bibitem{ShethTormen99}
R. K. Sheth and G. Tormen, \mnras {\bf 308}, 119 (1999).
\bibitem{ColeKaiser89}
S. Cole and N. Kaiser, \mnras {\bf 237} 1127 (1989).
\bibitem{MoWhite96}
H. J. Mo and S. D. M. White, \mnras {\bf 282}, 347 (1996).
\bibitem{TDS98} G. Tormen, A. Diaferio and D. Syer, \mnras {\bf 299}, 
728 (1998).
\bibitem{PD97} J. A. Peacock and S. J. Dodds, \mnras {\bf 280} L19 (1996).
\bibitem{Sigad00}
Y. Sigad, et al., to be submitted (2000).
\bibitem{LK} M. Kamionkowski and A. R. Liddle, preprint astro-ph/9911103 
(1999).
\bibitem{TE}
D. J. Eisenstein, W. Hu, J. Silk and A. S. Szalay, 
\apj {bf 494}, L1 (1998).
\bibitem{MWP99} A. Meiksin, M. White and J. A. Peacock, \mnras
{\bf 304}, 851 (1999).
\bibitem{White} M. White and R. A. C. Croft, preprint astro-ph/0001247 (2000).
\bibitem{Peebles} P. J. E. Peebles, {\it The Large Scale Structure 
of the Universe}, Princeton University Press (1980).
\bibitem{Jenkins} Jenkins, A. et al. \apj
{\bf 499} 20 (1998).
\bibitem{Diaferio99} 
A. Diaferio, G. Kauffmann, J. M. Colberg and S. D. M.
White, \mnras {\bf 307} 537 (1999).
\bibitem{zeldovich} Y. Zeldovich, \aa {\bf 5} 84 (1970). 
\bibitem{KNH} G. Kauffmann, A. Nusser and
M. Steinmetz, \mnras {\bf 286}, 795 (1997).
\bibitem{Carlberg} R. G. Carlberg et al., \apj {\bf 462} 32 (1996).
\bibitem{Guzik00} J. Guzik et al., in preparation (2000).
\end{thebibliography}
\end{document}